\documentclass[aip, amsmath, amssymb, reprint]{revtex4-2}

\usepackage{graphicx}
\usepackage{dcolumn}
\usepackage{bm}

\usepackage[utf8]{inputenc}
\usepackage[T1]{fontenc}
\usepackage{mathptmx}
\usepackage{etoolbox}

\usepackage[english]{babel}

\makeatletter
\def\@email#1#2{%
 \endgroup
 \patchcmd{\titleblock@produce}
  {\frontmatter@RRAPformat}
  {\frontmatter@RRAPformat{\produce@RRAP{*#1\href{mailto:#2}{#2}}}\frontmatter@RRAPformat}
  {}{}
}%
\makeatother
\begin{document}

\preprint{AIP/123-QED}

\title[Joule heating of a nanoemitter on the cathode surface]{Joule heating of an emitter on the cathode surface by field electron emission current with an account of the non-isolation of the apex}

\author{M. Diachenko}
\email{mykhailo.m.diachenko@gmail.com}

\author{S. Lebedynskyi}

\author{R. Kholodov}

\affiliation{Institute of Applied Physics, National Academy of Sciences of Ukraine, 58, Petropavlivska Street, Sumy, 40000, Ukraine}

\date{\today}

\begin{abstract}
This work is devoted to the investigation of the non-stationary problem of the thermal conductivity of a nanoemitter on the surface of a massive copper cathode when a field electron emission current passes through it. At the same time, the dependence of volume resistivity, thermal conductivity on temperature, and size effects have been taken into account. The influence of the Nottingham effect has been considered. The dependence of the equilibrium temperature of the emitter apex on the field enhancement factor for different values of the electric field strength has been found. Based on the assumption that the initial stage of the breakdown begins when the emitter apex melts, the conditions for the occurrence of a vacuum breakdown and the influence of the Nottingham effect have been analyzed. 
\end{abstract}

\maketitle

\section{\label{sec:level1} Introduction}

The physical nature of vacuum high-voltage breakdown, which can occur in elementary particle accelerators, in particular in CLIC (Compact LInear Collider, CERN), is quite complex and, despite numerous studies, a complete theory of this process still does not exist.

The electron emission from the cathode surface is usually local, due to the presence of emitters in the form of tips and other inhomogeneities on the surface. The presence of such objects leads to a local increase in the electric field strength, an increase in the field electron emission current density, and accordingly, heating of the emitter due to the dissipation of Joule energy. The process of resistive heating of the emitter on the cathode surface is an important stage of high-vacuum breakdown and is therefore relevant for theoretical research.

Due to the development of modern accelerators of charged particles, in which the gradients reach values at which high-vacuum breakdowns occur inside the accelerator structure, in particular at linear electron-positron colliders, many theoretical works dedicated to the study of these processes have recently appeared. Thus, in Refs.~\onlinecite{Djurabekova11, Parviainen11, Kyritsakis17, Eimre15, Kyritsakis18, Gao20, Mofakhami21}, the peculiarities of heating a nano-sized emitter on the surface of a massive cathode were investigated using molecular dynamics methods. Namely, in  Ref.~\onlinecite{Djurabekova11}, a molecular dynamic model was developed to describe the evolution of the emitter within the framework of the atomic molecular dynamics approach. In Ref.~\onlinecite{Parviainen11}, a model was developed to describe the evolution of a tip of a cylindrical shape caused by current, which is based on the methods of molecular dynamics, including resistive heating and electronic thermal conductivity. The consistency of the obtained simulation results with analytical expressions was also shown. In Ref.~\onlinecite{Kyritsakis17}, a method for calculating emission currents and the Nottingham effect was developed, while considering different modes of electron emission (thermal, field and intermediate). In Ref.~\onlinecite{Eimre15}, the dependence of the field electron emission current on the applied electric field was analyzed by the Fowler-Nordheim methods and the generalized formalism of the emission taking into account the temperature. It was shown that the usual Fowler–Nordheim equation leads to a significant underestimation of the electron emission current and can lead to a significant overestimation of the field enhancement factor, especially in the range of relatively low electric fields. In Ref.~\onlinecite{Kyritsakis18}, the processes taking place in a metal nanoemitter in the form of a truncated cone with an apex in the form of a hemisphere under conditions of intense electron emission were investigated. Multiscale simulations were also used, simultaneously including field-induced forces, electron emission, the Nottingham effect, and Joule heating. The possibility of evaporation of large fractions of the tip was shown, which can explain the origin of neutral atoms necessary for plasma initiation. In Ref.~\onlinecite{Gao20}, a combined method of molecular dynamics and electrodynamics was used to model the thermal evaporation of nanotips under the influence of a high electric field. In this paper, copper emitters with different initial geometries were investigated and as a result of simulations, it was shown that the aspect ratio of the emitters has a significant effect on thermal evaporation. In Ref.~\onlinecite{Mofakhami21}, the Nottingham effect was studied in detail for nano-sized emitters of a complex shape. It should also be noted Refs.~\onlinecite{Timko11, Timkophdthesis, Timko12, Bird94, Ejiri20} devoted to the simulation of interelectrode processes that occur at the initial stage of vacuum breakdown development. In Ref.~\onlinecite{Timko11}, a physical model was presented, based on the one-dimensional Particle In Cell (PIC) method, which describes the formation of plasma under the assumption that an emitter is initially present on the cathode, which enhancements the field. The model includes field electron emission and evaporation of neutral atoms as plasma initiation mechanisms and takes into account several surface processes and collisions between particles. As a result of the simulation, it was shown the possibility of plasma formation when the current density is $0.5-1~A/{\mu m}^2$, the ratio of neutral atoms to emission electrons is $0.01-0.05$ and time scales are $1-10~ns$. In Refs.~\onlinecite{Timkophdthesis, Timko12}, the previous model (Ref.~\onlinecite{Timko11}) was improved to a two-dimensional PIC model. Within the framework of these works, the temporal and spatial evolution of the vacuum arc was considered, which is created under the conditions of the existence of an emitter on the cathode surface. A current–voltage characteristic of the arc was qualitatively investigated and the conditions under which it occurs were considered. A comparison with the results of similar programs was also made and good consistency with them was shown. In Refs.~\onlinecite{Bird94, Ejiri20}, the two-dimensional PIC model was improved, taking into account collisions between particles using the Monte Carlo method, and more subtle effects were also considered. It should be noted the works of Refs.~\onlinecite{Tripathi20, Ludwick22, Jensen21, Jensen19}, in which both analytical and numerical studies of the current heating of carbon nanotubes are carried out. In particular, in Refs.~\onlinecite{Tripathi20, Ludwick22}, the spatial dependence of the temperature distribution along a carbon nanotube has been studied under various regimes of electron emission, taking into account the effects of Henderson-cooling and Nottingham-heating.

In this paper, within the framework of the theory of thermal conductivity, a model problem is considered, when there is a cylindrical nanoemitter on the cathode surface. At the same time, the Joule heating by the field electron emission current, the dependence of the resistivity on the temperature for a nano-sized tip, size effects, and the Nottingham effect are taken into account. It is shown that the process of evaporation of neutral atoms from the surface of the emitter can be neglected. An analysis of the conditions for the occurrence of a vacuum breakdown is carried out based on the assumption that it begins when the emitter apex melts, and the influence of the non-isolation of the tip on these conditions is investigated.

\section{\label{sec:level1} Theoretical background}

In this section, we will consider the theoretical basis of the heating of a copper emitter by the field electron emission current. We will consider a tip of cylindrical shape with a heat source in the form of Joule heating while taking into account the dependence of volume resistivity on temperature and size effects.

The nanoemitter will be located on a massive copper cathode. In the case of a cylindrical tip, we can switch to the one-dimensional case and the heat conduction equation has the following form
\begin{equation}
	\label{conduction_eq}
	c \rho \frac{{\partial T}}{{\partial t}} = \kappa \frac{{{\partial ^2} T}}{{\partial {x^2}}} + q\left( T \right),
\end{equation}
where $c$  is the specific heat capacity, $\rho$ is the density of the tip material, $\kappa$ is the thermal conductivity.

In Eq.~(\ref{conduction_eq}), $q$ determines the power of heat sources, and in the case of Joule heating by current, the expression for $q$ has the form
\begin{equation}
	\label{q}
	q\left( T \right) = {\rho_r}\left( T \right){j^2}.
\end{equation}
where $\rho_r$ is the resistivity, $j$ is the current density.

In the problem of thermal conductivity of nano-sized tips of cylindrical shape, when the diameter is smaller than the average length of the free path of electrons, it is also necessary to take into account size effects. These effects significantly affect the resistivity of a substance and the thermal conductivity. In accordance with  Ref.~\onlinecite{Parviainen11}, the expression for the resistivity can be written in the form
\begin{equation}
	\label{size_effect}
	{\rho_r} = \frac{\eta}{r}\frac{{{\rho _0}}}{{{T_0}}}T,
\end{equation}
where  $\rho_0$ is the value of resistivity at a temperature $T_0$, $r$ is the radius of the cylinder, $\eta = 70~nm$. This expression is obtained from the results of computer simulation and within the values of $0.5~nm$ $<r<10 ~nm$ the error is less than 6\%.

The size effects also affect the coefficient of thermal conductivity, and according to the Wiedemann-Franz-Lorenz law, it can be written as
\begin{equation}
	\label{q}
	\kappa  = \frac{{LT}}{\rho } = \frac{{rL{T_0}}}{{\eta{\rho _0}}},
\end{equation}
where $L$ is the Lorenz number.

Note that there are more sophisticated models of the electrical resistivity and the thermal conductivity (Refs. 1 and 2), but in this work, to simplify the numerical solution algorithm of the non-stationary thermal conductivity equation, we limit ourselves to the above dependencies.

The current density of the field emission from the emitter on the metal surface due to electron tunneling through the potential barrier at the metal-vacuum interface is described by a Murphy-Good model (Ref.~\onlinecite{Murphy56}):
\begin{equation}
	\label{Fouler_eq}
	{j} = \lambda_T \frac{a}{{{t^2}\left( y \right)}}\frac{{{F^2}}}{\varphi }\exp \left( { - b\frac{{{\varphi ^{{3 \mathord{\left/
	{\vphantom {3 2}} \right. \kern-\nulldelimiterspace} 2}}}}}{F}\vartheta \left( y \right)} \right),
\end{equation}
where $\lambda_T$ is a temperature correction factor 
\begin{equation*}
	\lambda_T = \frac{\pi k s T}{\sin (\pi k s T)},
\end{equation*}
and the notation is entered
\begin{equation*}
	s=\frac{3b\sqrt{\varphi}t(y)}{2F},~~y = d \frac{{\sqrt F }}{\varphi },~~a = \frac{{{e^3}}}{{16{\pi ^2}\hbar }},~~b = \frac{{4\sqrt {2m} }}{{3\hbar e}},
\end{equation*}
$ d\approx 3.79 \cdot 10^{ - 4} eV {cm}^{1/2} {V}^{-1/2}$.

Eq.~(\ref{Fouler_eq}) also takes into account the local enhancement of the electric field at the emitter apex, and thus the local field $F = \beta E$, where $E$ is the macroscopic field and $\beta$ is the enhancement factor. If the emitter has the shape of a cylinder, then it is approximately assumed that this factor is determined by the ratio of the height to the radius of the emitter $\beta = h/r$. Henceforth, we will use this approximation for the coefficient $\beta$.  

For numerical calculations, the functions $t(y)$ and $\vartheta (y)$ can be approximately taken in the following form (Ref.~\onlinecite{Forbes06})
\begin{equation}
\label{Forbes}
{t}\left( y \right) \approx 1+\frac{y^2}{9}\left(1-lny\right),~~\vartheta \left( y \right) \approx 1 - y^2 + \frac{y^2}{3} lny.
\end{equation}

The error with this choice of functions (\ref{Forbes}) is less than 0.6\% for values of $y$ in the range from 0 to 1. In the numerical calculations conducted in this article, parameter y has a maximum value of 0.78, which is within the relevant range of values.

It should be emphasized that the Fowler-Nordheim formula is obtained for the case of zero temperature of the metal, but in real conditions, the temperature differs from zero due to the heating of the emitter by the field electron emission current, so the effect of temperature on the process under study should also be taken into account. But within the framework of this work, we will neglect this dependence.

To solve Eq.~(\ref{conduction_eq}), it is also necessary to specify the initial and boundary conditions. In this work, the initial temperature of the emitter will be $T_0 = 293.15~K$, a constant temperature will be maintained on the lower base of the tip (the point of contact with the cathode), which coincides with the initial temperature $T_0$. For the upper base, we will consider two cases, when it is isolated (there is no heat flow through the top base of the tip) and when it is not isolated due to the sublimation process and the Nottingham effect. For the first case, the initial and boundary conditions of the problem are as follows
\begin{equation}
	\label{limit_con}
	{\left. T \right|_{t = 0}} = {T_0},~~{\left. T \right|_{x = 0}} = {T_0},~~{\left. {\frac{{\partial T}}{{\partial x}}} \right|_{x = h}} = 0.
\end{equation}

Let us first consider the stationary case, when the change in temperature over time can be neglected, then the heat conduction equation Eq.~(\ref{conduction_eq}) can be rewritten in the form
\begin{equation}
	\label{conduction_stationary_eq}
	\frac{{{\partial ^2} T}}{{\partial {x^2}}} + \zeta T = 0,
\end{equation}
where $\zeta = q/(\kappa T)$. In this case, the boundary conditions have the form Eq.~(\ref{limit_con}).

The temperature distribution can be found from the equation (\ref{conduction_stationary_eq}) in the form
\begin{equation}
	\label{temperature_x}
	T(x) = {T_0}\left[ {\cos (\sqrt \zeta x) + tg(\sqrt \zeta h)\sin (\sqrt \zeta x)} \right].
\end{equation}
From Eq.~(\ref{temperature_x}), the temperature on the upper base of the cylinder is determined as
\begin{equation}
	\label{temperature_appex}
	T|_{x=h} = \frac{{{T_0}}}{{\cos (\sqrt \zeta h)}}.
\end{equation}
Taking into account the formula (\ref{temperature_appex}), we can obtain an expression for the current density at which melting of the tip occurs. To do this, it is necessary to substitute the melting temperature $T_m$ into the left-hand side of Eq. (10) and solve it for the current density. This way, we can obtain  
\begin{equation}
	\label{current_melting}
	j_m = \frac{{{T_0}\sqrt L }}{{\eta{\rho _0}\beta }}\arccos \frac{{{T_0}}}{{{T_m}}},
\end{equation}
where $T_m$ is the melting point, $j_m$ is the current density at which the tip temperature reaches the melting point.

\section{\label{sec:level1} Numerical calculations}

In this section, we will consider the non-stationary problem of the thermal conductivity of a cylindrical tip during Joule heating by a current with density (\ref{Fouler_eq}). We will numerically solve the equation (\ref{conduction_eq}) using the implicit Euler scheme, as well as the Thomas algorithm. To do this, we will reduce the equation of thermal conductivity to a dimensionless form by introducing the following dimensionless quantities
\begin{equation}
	\label{dimensionless_quantities}
	\tilde T = \frac{T}{{{T_0}}},~~\tilde x = \frac{x}{h},~~\tilde t = \frac{t}{{{t_0}}},~~{t_0} = \frac{{\eta{\rho _0}c\rho {h^2}}}{{rL{T_0}}}.
\end{equation}
Considering the introduced dimensionless quantities (\ref{dimensionless_quantities}), the thermal conductivity equation can be rewritten in the following form
\begin{equation}
	\label{conduction_eq_2}
	\frac{{\partial \tilde T}}{{\partial \tilde t}} = \frac{{{\partial ^2} \tilde T}}{{\partial {\tilde x^2}}} + Q\tilde T,
\end{equation}
where 
\begin{equation*}
	Q = \frac{1}{L}{\left( {\frac{{\eta{\rho _0}}}{{{T_0}}}\beta j} \right)^2}.
\end{equation*}

\subsection{\label{sec:level2} Case of an isolated tip apex}

First, let us consider the case when the emitter apex is isolated. We will consider the initial and boundary conditions in the form (\ref{limit_con}). At the same time, these conditions can be rewritten in terms of dimensionless quantities as follows
\begin{equation}
	\label{limit_con_2}
	{\left. \tilde T \right|_{\tilde t = 0}} = 1,~~{\left. \tilde T \right|_{\tilde x = 0}} = 1,~~{\left. {\frac{{\partial \tilde T}}{{\partial \tilde x}}} \right|_{\tilde x = 1}} = 0.
\end{equation}

Using Euler's implicit scheme, the heat conduction equation can be written
\begin{equation}
	\label{Euler_scheme}
	\frac{{\tilde T_i^{n + 1} - \tilde T_i^n}}{{\delta \tilde t}} = \frac{{\tilde T_{i + 1}^{n + 1} - 2\tilde T_i^{n + 1} + \tilde T_{i - 1}^{n + 1}}}{{\delta {{\tilde x}^2}}} + Q\tilde T_i^{n + 1},
\end{equation}
where $\delta \tilde x$, $\delta \tilde t$ are steps by coordinate and time, index $n$ corresponds to the grid node number by time, and $i$ by coordinate.

The equation (\ref{Euler_scheme}) can be rewritten in a more convenient form for applying the Thomas algorithm
\begin{equation}
	\label{Euler_scheme_2}
	\tilde T_{i + 1}^{n + 1} - {B_i}\tilde T_i^{n + 1} + \tilde T_{i - 1}^{n + 1} = {G_i}.
\end{equation}
The expression (\ref{Euler_scheme_2}) has the following notation
\begin{equation*}
	{G_i} =  - \frac{{\delta {{\tilde x}^2}}}{{\delta \tilde t}}\tilde T_i^n,~~{B_i} = 2 + \frac{{\delta {{\tilde x}^2}}}{{\delta \tilde t}} - Q\delta {\tilde x^2}.
\end{equation*}

As part of the Thomas algorithm, it is assumed that there are such sets of numbers $\alpha_i$ and $\xi_i$ for which the equality holds
\begin{equation}
	\label{Thomas_scheme}
	\tilde T_i^{n + 1} = {\alpha _i}\tilde T_{i + 1}^{n + 1} + {\xi_i}.
\end{equation}
Given the expression (\ref{Euler_scheme_2}), we can find recurrence relations for these coefficients
\begin{equation}
	\label{Thomas_scheme_recurrent}
	{\alpha _i} = \frac{1}{{{B_i} - {\alpha _{i - 1}}}},~~{\xi_i} = \frac{{{\xi_{i - 1}} - {G_i}}}{{{B_i} - {\alpha _{i - 1}}}}.
\end{equation}
To determine the coefficients $\alpha_i$, $\xi_i$ (\ref{Thomas_scheme_recurrent}), it is necessary to find $\alpha_0$ and $\xi_0$ , which can be obtained from the lower boundary condition
\begin{equation}
	\label{Thomas_scheme_recurrent_0}
	{\alpha _0} = 0,~~{\xi_0} = 1.
\end{equation}
Since the calculation scheme has the second order of accuracy in terms of $\delta \tilde x$, it is necessary to discretize the upper boundary condition with $O\left( {\delta {{\tilde x}^2}} \right)$ accuracy. For this, it is necessary to expand the function $\tilde T (x)$ into a Taylor series around the point $\tilde x = 1$ to terms of the second order with respect to $\delta \tilde x$
\begin{equation}
	\label{series}
	\tilde T_{N - 2}^{n + 1} = \tilde T_{N - 1}^{n + 1} - \delta \tilde x\left. {\frac{{\partial \tilde T}}{{\partial \tilde x}}} \right|_{N-1}^{n + 1} + \frac{{\delta {{\tilde x}^2}}}{2}\left. {\frac{{{\partial ^2}\tilde T}}{{\partial {{\tilde x}^2}}}} \right|_{N-1}^{n + 1}.
\end{equation}
Taking into account the upper boundary condition, that is, the isolation of the upper base of the tip, and the heat conduction equation (\ref{conduction_eq_2}), the following relation can be written 
\begin{equation}
	\label{series_2}
	\tilde T_{N - 2}^{n + 1} = \tilde T_{N - 1}^{n + 1} + \frac{{\delta {{\tilde x}^2}}}{2}\left( {\frac{{\tilde T_{N - 1}^{n + 1} - \tilde T_{N - 1}^n}}{{\delta \tilde t}} - Q\tilde T_{N - 1}^{n + 1}} \right).
\end{equation}
Thus, the temperature at the coordinate grid node $N-2$ at the moment $n+1$ according to (\ref{series_2}) is determined by the temperature value at the last node $N-1$. The expression for $\tilde T_{N - 2}^{n + 1}$ can also be written in terms of Thomas algorithm coefficients as follows
\begin{equation}
	\label{series_3}
	\tilde T_{N - 2}^{n + 1} = {\alpha _{N - 2}}T_{N - 1}^{n + 1} + {\xi_{N - 2}}.
\end{equation}
Considering the expressions (\ref{series_2}) and (\ref{series_3}), we can find
\begin{equation}
	\label{series_4}
	\tilde T_{N - 1}^{n + 1} = \frac{{{\xi_{N - 2}} + \frac{{\delta {{\tilde x}^2}}}{{2\delta \tilde t}}\tilde T_{N - 1}^n}}{{1 - {\alpha_{N - 2}} + \frac{{\delta {{\tilde x}^2}}}{{2\delta \tilde t}} - \frac{{\delta {{\tilde x}^2}}}{2}Q}}.
\end{equation}
In this way, the solution of the heat conduction equation (\ref{conduction_eq_2}) by the Thomas algorithm is reduced to the following stages. First, the coefficients $\alpha_0$ and $\xi_0$ are calculated from the lower boundary condition, then $\alpha_i$ and $\xi_i$ are found by direct running ($i= 1,~2,~...~,~N - 1 $), at the last stage the temperatures in the other nodes are found $T_i ^{n + 1}$ ($i= N-2,~N-3,~...~,~1$) by the inverse run according to the formula (\ref{Thomas_scheme}). This is what one iteration looks like over time.

First, for numerical calculations according to the above scheme, we will choose the following parameters of the problem $r=2.2~nm$, $h=100~nm$, $E=170~MV/m$, the initial and boundary conditions will be (\ref{limit_con}). In this paper, we will also consider an emitter made of oxygen-free electronic copper (C10100), the parameters of which are given in Ref.~\onlinecite{Davis01}. It should be noted that for these calculations, we do not take into account the dependence of the current density on the temperature in Eq.~(5), since the main task was to compare the numerical results with the analytical solution for the stationary case. However, the temperature effect is taken into account in further calculations. Figure 1 shows the results of the numerical calculation, from which it can be seen that for the given parameters, the greatest heating is observed on the tip apex and the temperature distribution becomes unchanged over time. Figure 2 shows the dependence of the temperature of the emitter apex on the time. As can be seen from this figure, after a time of the order of $t_0$, the temperature reaches a stationary mode and coincides with the analytical value (\ref{temperature_appex}) for the stationary case. It should be noted that for the given calculation parameters, $t_0 \approx 2.8~ns$. Figure 3 shows the dependence of the temperature on the coordinate after the onset of the stationary mode ($t=30 t_0$). This figure also shows a good agreement between numerical calculations and analytical expressions.
\begin{figure}
	\includegraphics[width=1\linewidth]{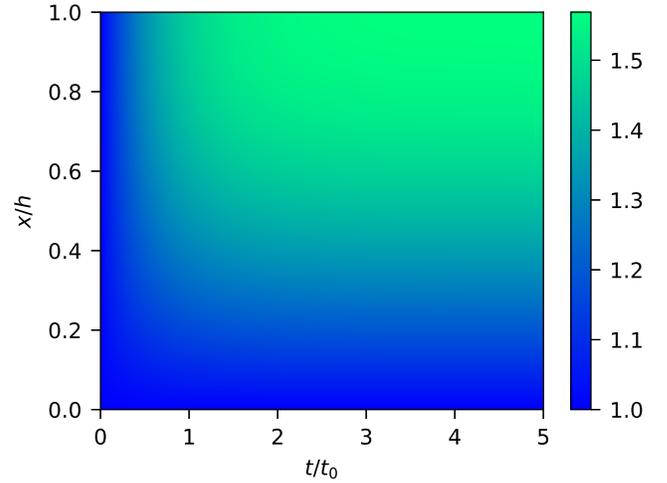}
	\caption{\label{colormap} Dependence of tip temperature on coordinate and time for the case of an isolated apex.}
\end{figure}
\begin{figure}
	\includegraphics[width=1\linewidth]{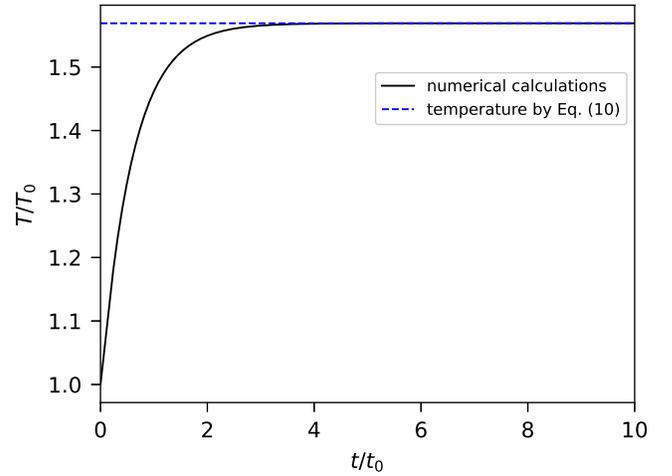}
	\caption{\label{T_t} Dependence of the temperature of the tip apex on time.}
\end{figure}
\begin{figure}
	\includegraphics[width=1\linewidth]{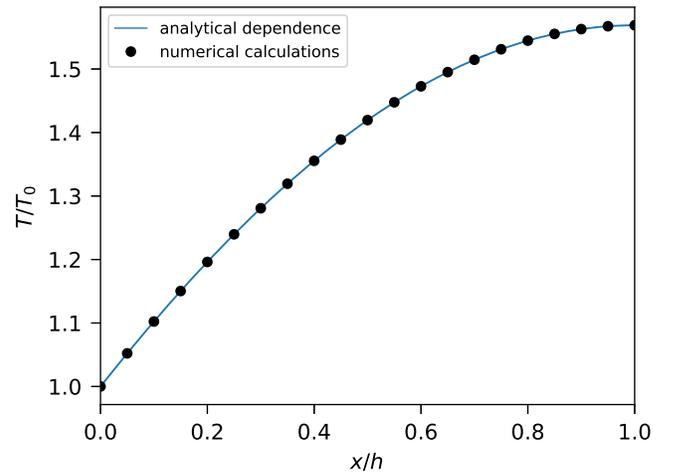}
	\caption{\label{T_x} Temperature distribution along the coordinate after heating during time $10 t_0$.}
\end{figure}

The dependence of the temperature of the emitter apex on the field enhancement factor $\beta$, that is, on the radius of the tip at a fixed height, for different values of the electric field strength was also considered. The results of numerical calculations are demonstrated in figure 4. If it is assumed that the vacuum breakdown begins when the emitter apex reaches the melting point, then the $\beta$ coefficient when the breakdown begins can be found for different values of the electric field. Figure 6 shows such a dependence, and as we can see, for an electric field strength of $100~MV/m$, the enhancement factor is 76.3, at lower values, breakdown does not occur. If the field strength is $170~MV/m$, then the coefficient is 47.4, which is consistent with experimental results (Ref.~\onlinecite{Descoeudres09}), in which it was approximately equal to 50 for a given value of the electric field. From the interpolation given in the figure 4, it is also possible to determine $\beta$ for other values of the electric field.
\begin{figure}
	\includegraphics[width=1\linewidth]{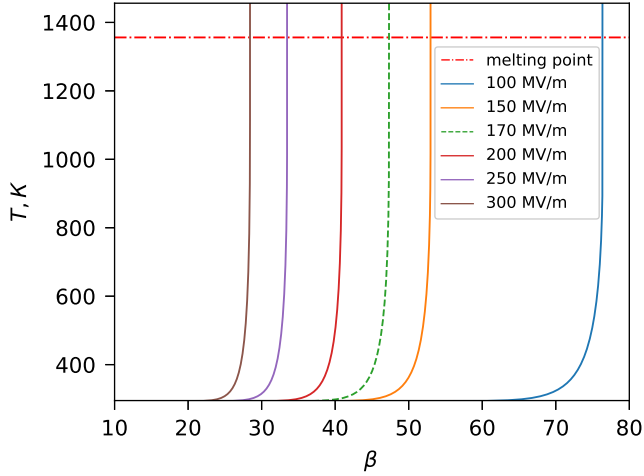}
	\caption{\label{T_beta} Dependence of the equilibrium temperature of the emitter apex on $\beta$ at different values of the applied electric field.}
\end{figure}

\subsection{\label{sec:level2} Case of a non-isolated tip apex}

In previous calculations, we used the approximation that the emitter is isolated. Henceforth, we will take into account the non-isolation of its apex due to the Nottingham effect. We will show that the process of sublimation can be neglected under the investigated conditions. To solve the heat conduction equation, we will use the computational scheme as in the previous subsection, but to take into account the boundary condition on the upper base of the tip, we will also use the method of successive iterations.

\subsubsection{\label{sec:level3} Sublimation}

To evaluate the influence of the sublimation process, we will calculate the mass of the substance evaporated from the tip at the melting temperature. We will use the Hertz-Knudsen equation, according to which the rate of sublimation of neutral atoms from the cathode surface is determined as follows
\begin{equation}
	\label{sublimation_velocity}
	{w_{evap}} = \sqrt {\frac{M}{{2\pi RT}}} p,
\end{equation}
where $M$ is the molar mass of the substance, $R$ is the universal gas constant, and $p$ is the saturated vapor pressure, which is a function of temperature and is determined from an empirical formula that looks like this (Ref.~\onlinecite{ Alcock84})
\begin{equation}
	\label{empirical_pressure}
	\lg \tilde p = A + \frac{B}{{\tilde T}} + C\lg \tilde T + D\tilde T \cdot {10^{ - 3}}.
\end{equation}
It should be noted that in this paper we consider the process of heating the emitter before its melting, so in the expression (\ref{empirical_pressure}) we will use empirical parameters for the solid phase of the substance, which take the values (Ref.~\onlinecite{Alcock84}) $A = 7.810, ~B = - 17687, ~C = - 0.2638, ~D = - 0.1486$.

Let us calculate the mass of the substance that evaporates as a result of the sublimation process during the time $10 t_0$. Taking into account the expression (\ref{empirical_pressure}), the rate of this process is $w_{evap} \approx 5\cdot 10^{-5}~kg/m^2 s$. Then $m_{evap} \approx 1.4\cdot 10^{-28}~kg$, which is much smaller than the mass of a copper atom. Thus, the process of sublimation can be neglected when we consider the process of heating the tip before the start of melting. Henceforth, the evaporation of neutral atoms will not be considered in numerical calculations.  

\subsubsection{\label{sec:level3} Nottingham effect}

In this subsection, we will take into account the Nottingham effect. It consists of the difference in the average energy of electrons that come from the depth of the cathode to its surface and electrons that leave it. At the same time, if the temperature of the cathode surface is lower than the inversion temperature, then the average energy of the electrons leaving the surface will be lower than the energy of the electrons coming from the depth of the metal, and heating of the tip apex will be observed (the Nottingham heating). If the temperature is greater than the inversion temperature, Henderson cooling will be observed on the contrary. The inversion temperature is determined by the expression
\begin{equation}
	\label{temperature_inversion}
	{T_{inver}} = 5.67 \cdot {10^{ - 5}}\frac{{\beta E}}{{\sqrt \varphi  }},
\end{equation}
where $E$ is the electric field strength in $V/cm$, $\varphi$ is the work function of the electron in $eV$, $T_{inver}$ is the inversion temperature in $K$.

Let us estimate the temperature (\ref{temperature_inversion}) for the parameters of the problem. Thus, for the value of the electric field $E = 170~MV/m$, the enhancement factor $\beta=50$ and the work function of the electron  $\varphi=4.5~eV$, we have that ${T_{inver}} \approx 2272~K$. As we can see, the inversion temperature is much higher than the melting temperature of copper, so we will consider only the case when the Nottingham heating occurs. 

For the case when ${T_{emis}} < 1.2 \cdot {T_{inver}}$, the difference in the average energy of electrons leaving the metal surface and electrons coming from the depth of the cathode has the form (Ref.~\onlinecite {Litvinov83})
\begin{equation}
	\label{Nottingem_energy}
	\Delta \varepsilon  = \frac{{{\pi ^2}}}{2}{\left( {\frac{{k{T_{emis}}}}{{{\varepsilon _F}}}} \right)^2}{\varepsilon _F} + 2k\pi T_{emis} ctg\frac{{\pi {T_{emis}}}}{{2{T_{inver}}}},
\end{equation}
where $\varepsilon _F$ is the Fermi energy.

Considering the Nottingham effect, the boundary condition on the upper base of the emitter has the following form
\begin{equation}
	\label{Nottingem_limit}
	{\left. {\frac{{\partial \tilde T}}{{\partial \tilde x}}} \right|_{\tilde x = 1}} = \frac{j}{e}\Delta \varepsilon \frac{h}{{\kappa {T_0}}},
\end{equation}
where $e$ is the electron charge.

In the second approximation by a step along the coordinate grid, from the boundary condition (\ref{Nottingem_limit}), the relation can be found
\begin{equation}
	\label{Nottingem_limit_2}
	\tilde T_{N - 1}^{n + 1} = \frac{{{\xi_{N - 2}} + \frac{{\delta {{\tilde x}^2}}}{{2\delta \tilde t}}\tilde T_{N - 1}^n + V\Delta \varepsilon (\tilde T_{N - 1}^{n + 1})\delta \tilde x}}{{1 - {\alpha_{N - 2}} + \frac{{\delta {{\tilde x}^2}}}{{2\delta \tilde t}} - Q\frac{{\delta {{\tilde x}^2}}}{2}}}.
\end{equation}
Figures 5 and 6 present the results of numerical calculations of the heat conduction equation, taking into account the Nottingham effect. As can be seen from Fig.~5, the curves have a smaller slope than in the case of an isolated tip, and the points of intersection of the curves with the melting temperature are shifted to the left. This can also be seen in Fig.~6. If we assume that the initial stage of the breakdown occurs when the tip apex begins to melt, then taking into account the Nottingham effect leads to a decrease in the breakdown electric field, which can be seen in Fig.~6. For example, if $\beta = 50$, the reduction is approximately 10\%.
\begin{figure}
	\includegraphics[width=1\linewidth]{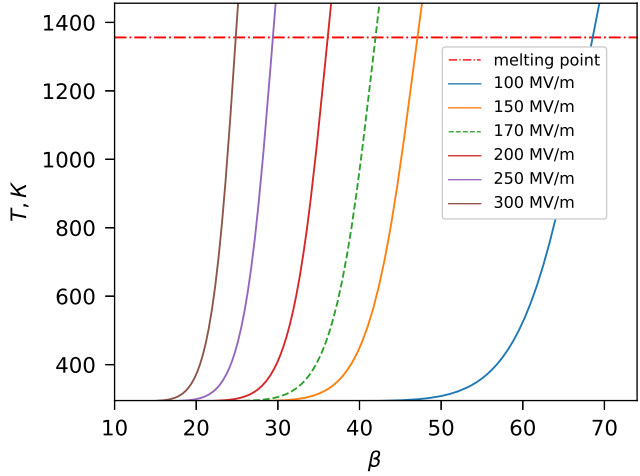}
	\caption{\label{T_beta_N} Temperature distribution along the coordinate after heating during time $30 t_0$, taking into account the Nottingham effect.}
\end{figure}
\begin{figure}
	\includegraphics[width=1\linewidth]{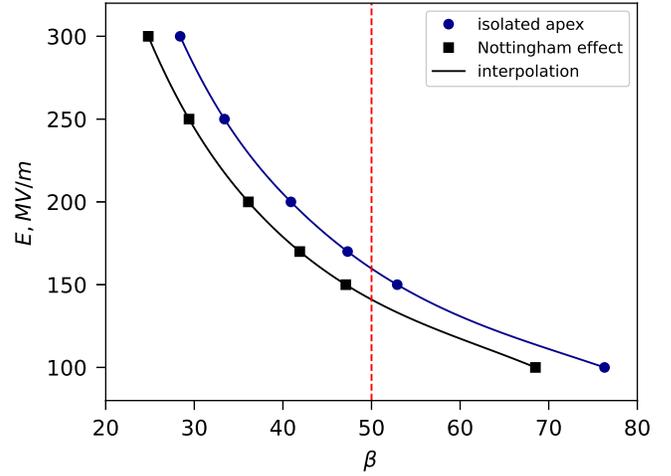}
	\caption{\label{E_beta}Dependence of the breakdown electric field strength on the field enhancement factor $\beta$ for the cases of isolated emitter apex and in consideration of the Nottingham heating.}
\end{figure}

\section{\label{sec:level1} Conclusions}

In this paper, the non-stationary problem of thermal conductivity during Joule heating by a current of a cylindrical nanoemitter, which is located on a massive copper cathode, was considered. At the same time, the dependence of resistivity and thermal conductivity on temperature, as well as size effects were taken into account.

The dependence of the temperature of the emitter apex on the field enhancement factor for different values of the electric field strength was obtained. The conditions for the occurrence of a vacuum breakdown are analyzed based on the assumption that it begins when the emitter apex melts. It is shown for the case of an isolated copper emitter that at an electric field strength of $100~MV/m$, the enhancement factor is $76.3$, which corresponds to a radius of $1.3~nm$ and a height of $100~nm$ at larger values of the radius, breakdown does not occur. If $E = 170~MV/m$, then this factor is $47.3$, which is consistent with both experimental and theoretical results (Ref.~\onlinecite{Parviainen11}), in which this factor was approximately equal to $50$ for a given value of the electric field.

It is shown that the sublimation of neutral atoms until the moment of melting of the emitter apex can be neglected. It is numerically shown that taking into account the Nottingham effect leads to a decrease in the breakdown electric field strength, so for $\beta = 50$, the decrease is approximately $12\%$.

\begin{acknowledgments}
	The publication is based on the research provided by the grant support of the National Academy of Sciences of Ukraine (NASU) for research laboratories/groups of young scientists of the National Academy of Sciences of Ukraine to research priority areas of development of science and technology in 2021–2022 under contract No 16/01-2021 (3).
\end{acknowledgments}

\section*{Author Declarations}
\subsection*{Conflict of Interest}
The authors have no conflicts to disclose.

\section*{Data Availability}
The data that support the findings of this study are available from the corresponding author upon reasonable request.

\bibliography{aipsamp}

\end{document}